\documentclass[twocolumn]{IEEEtran} 
\usepackage{graphicx}
\usepackage{epsf}       
\usepackage{epsfig}       
\usepackage{amssymb}

\def\BibTeX{{\rm B\kern-.05em{\sc i\kern-.025em b}\kern-.08em
    T\kern-.1667em\lower.7ex\hbox{E}\kern-.125emX}}

\begin{document}

\title{A Fast High Resolution Track Trigger\\ for the H1 Experiment}

\author{ 
  \thanks{Manuscript received November 3, 2000.} 
  \thanks{Dortmund is supported by the BMBF; grant\,no.\,057DO55P.}  
  A.~Baird\thanks{A.~Baird, CLRC Rutherford Appleton Lab., Oxfordshire, UK},
  E.~Elsen\thanks{E.~Elsen, DESY, Hamburg, Germany}, 
  Y.~H.~Fleming\thanks{Y.~H.~Fleming, University of Birmingham, Birmingham, UK},
  M.~Kolander\thanks{M.~Kolander, Universit\"at Dortmund, Dortmund, Germany}, 
  S.~Kolya\thanks{S.~Koyla, University of Manchester, Manchester, UK},
  D.~Meer\thanks{D.~Meer, ETH Z\"urich, Z\"urich, Switzerland}, 
  D.~Mercer\thanks{D.~Mercer, University of Manchester, Manchester, UK},
  J.~Naumann\thanks{J.~Naumann, Universit\"at Dortmund, Dortmund, Germany}, 
  P.~R.~Newman\thanks{P.~R.~Newman, University of Birmingham, Birmingham, UK},
  D.~Sankey\thanks{D.~Sankey, CLRC Rutherford Appleton Lab., Oxfordshire, UK}, 
  A.~Sch\"oning\thanks{A.~Sch\"oning, ETH Z\"urich, Z\"urich, Switzerland},
  H.-C.~Schultz-Coulon\thanks{H.-C.~Schultz-Coulon, Universit\"at Dortmund, Dortmund, Germany}, 
  Ch.~Wissing\thanks{Ch.~Wissing, Universit\"at Dortmund, Dortmund, Germany}
}

\markboth{IEEE Transactions On Nuclear Science, Vol. XX, No. Y, Month 2000}
{A Fast High Resolution Track Trigger for the H1 Experiment} 

\maketitle

\begin{abstract}
  After 2001 the upgraded \boldmath{$ep$} collider HERA will provide an about
  five times higher luminosity for the two experiments H1 and ZEUS.  In order to
  cope with the expected higher event rates the H1 collaboration is building a
  track based trigger system, the Fast Track Trigger (FTT). It will be
  integrated in the first three levels (L1--L3) of the H1 trigger scheme to
  provide higher selectivity for events with charged particles. The FTT will
  allow to reconstruct 3-dimensional tracks in the central drift chamber down to
  100 MeV/c within the L2 latency of \boldmath{$\sim$}23~\boldmath{$\mu$}s.  To
  reach the necessary momentum resolution of $\sim$5\% (at 1~GeV/c)
  sophisticated reconstruction algorithms have to be implemented using high
  density Field Programmable Gate Arrays (FPGA) and their embedded Content
  Addressable Memories (CAM). The final track parameter optimization will be
  done using non-iterative fits implemented in DSPs.
  While at the first trigger level rough track information will be provided, at
  L2 tracks with high resolution are available to form trigger decisions on
  topological and other track based criteria like multiplicities and momenta.
  At the third trigger level a farm of commercial processor boards will be used
  to compute physics quantities such as invariant masses.
\end{abstract}


\section{Introduction}
\PARstart{W}{ith} the HERA collider at DESY~920 GeV protons are collided with
27.6~GeV electrons (positrons) every 96~ns. At one of the interaction points the
H1 detector is located, an omni-purpose detector described in \cite{cite:h1}.
One of its main components is the central jet chamber (CJC) which consists of
two concentric drift chambers, the inner CJC1 and the outer CJC2, with 24 and 32
layers of wires, respectively. Its signals will be the basic input for the high
resolution track trigger described in this paper.

H1 uses a four level multi-stage trigger system (L1-L4) of which the first level
\cite{cite:l1} is a fully pipelined, dead time free hardware trigger with a
decision time of 2.3~$\mu$s, reducing the 10 MHz input rate by more than four
orders of magnitude. The L1 decision is refined by L2 \cite{cite:l2} within a
fixed time of 23~$\mu$s. Subsequently the detector read-out starts which in
total takes about 1 ms but can be aborted by a negative decision generated in
parallel by trigger level~3. To have a beneficial effect on the dead time the
L3 decision should be available not more than 100~$\mu$s after the read-out has
started.

After an event has successfully been read out at the front end pipelines are
restarted. Events are collected via an optical fiber ring, sent and buffered at
a filter farm (L4) where they are fully reconstructed within some 100~ms. A
detailed description of the H1 data acquisition system can be found in
\cite{cite:daq}.

To study rare interactions with high precision, the HERA ring will be upgraded
providing about a factor of five larger luminosity from 2001 onwards. The
increased $ep$ and background rates will however place a considerable strain on
the H1 data acquisition. This necessitates an upgrade of the existing trigger
system to provide better selectivity for the many exclusive final states
exhibiting characteristic track based topologies -- especially those containing
heavy quarks -- with rather small cross sections and low transverse momentum
$p_t$ final state particles. 

In contrast, events with a high momentum transfer $Q^2 > $ 100 GeV$^2$ can
efficiently be triggered by calorimeter based signals with the present system
also after the upgrade. No such possibility exists for events at lower $Q^2$ and
additional selection power is needed. Hence the H1 collaboration is building a
Fast Track Trigger (FTT) which is able to find and reconstruct tracks and
particle resonances at the first three trigger levels and provides trigger
signals derived from track based quantities. The development of such a track
trigger performing these sophisticated tasks within the tight time constraints
given by the H1 trigger system has only become possible with the recent
improvements in size and speed of available electronic components.

\section{The Fast Track Trigger}
\begin{figure}[b]
\centerline{\includegraphics[width=3.5in,clip]{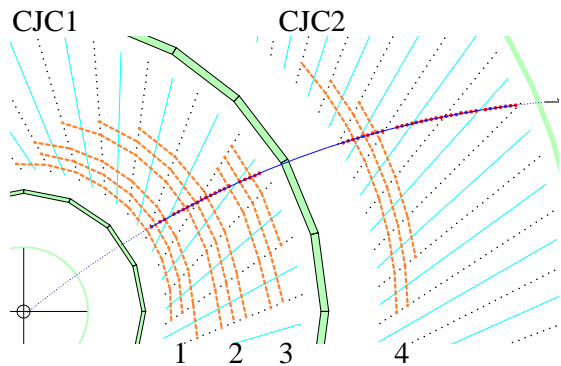}}
\caption{$r-\phi$ view of a charged particle track from the interaction point 
  traversing the central drift chamber of the H1 experiment. In addition to the
  boundaries of the chambers the sense and cathode wires are indicated. The four
  trigger layers formed out of three layers of wires each are marked by the
  thick dashed lines.}
\label{figure:layers}
\end{figure}

The FTT functionality is based on CJC information derived from four groups of
three layers of wires each, three of them inside CJC1 and one inside CJC2 as
shown in figure~\ref{figure:layers}, which also displays the geometrical cell
structure of both chambers. It is designed to handle up to 48 tracks which is
sufficient for about 98\% of the events of interest.

After digitizing the drift chamber signals a search for track segments within a
group of three drift chamber hits is performed in each of the four radial
trigger layers. A subsequent coarse track segment linking is completed within
2.1~$\mu$s to contribute to a level 1 trigger decision based on track
multiplicities and coarse $p_t$ cuts.  A positive L1 decision triggers a refined
track segment search reusing the FTT L1 hardware. The result is used by the
second level FTT where the track segments have to be linked and fitted within 20
$\mu$s including the determination of event quantities like a refined track
multiplicity, momentum sums and invariant masses for low multiplicity events.
The track parameters of the fitted tracks are sent to FTT L3 where a full search
for particle resonances is performed within 100~$\mu$s and the L3 track
information is either used directly or in combination with information from
other trigger subsystems to generate a final L3 decision.
 
\begin{figure}[tb]
\centerline{\includegraphics[width=3.5in,clip]{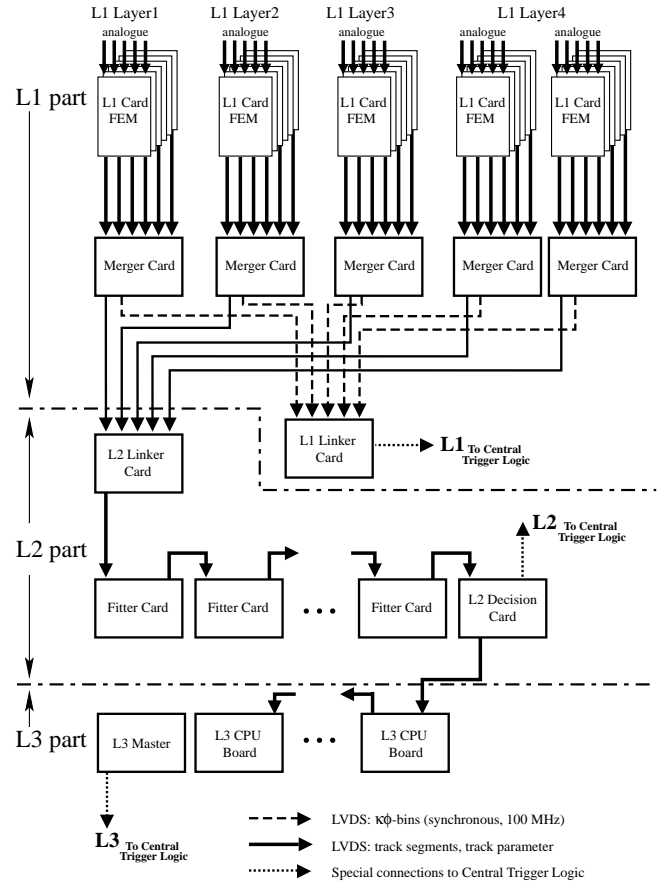}}
\caption{The hardware realization of the FTT. After signal
  digitization and hit recognition the track segment finding is done on the
  Front End Modules (FEM). Via LVDS links and intermediate Merger Cards the
  track segments are first fed to the L1 Linker Card to generate a level~1
  trigger decision and after a refined track segment finding sent to the L2
  Linker Card, where full tracks are extracted followed by a 3-dimensional
  fitting on a Fitter Card. A Decision Card calculates second level trigger
  signals and serves as a link to the third level trigger where invariant mass
  sums are calculated on CPU boards. Each CPU board is used for a specific
  physics channel.}
\label{figure:ftt-schema}
\end{figure}

The different components of the FTT are schematically shown in
figure~\ref{figure:ftt-schema} and will be described in detail below.

\subsection{Hit and Track Segment Finding at L1}
Adapter cards will be used to tap selected analog CJC signals from the
existing drift chamber readout. At the analog part of the so called Front End
Modules (FEM) the signals of both wire ends are digitized at 80~MHz using a
common 8 bit linear dual FADC (AD9288), where 15 FADCs are mounted on one board
to serve 5 neighboring drift chamber cells. The main part of the following
digital signal processing is done by 5 Altera APEX 20K600E FPGAs
\cite{cite:altera}, one for each group of wires. Hit finding is performed by
looking for pulses exceeding the noise level and extracting their precise time
information with a precision of 2-3 ns. For each hit the $z$-coordinate is
obtained by a charge devision technique based on the signals from both wire
ends. A resolution of $\sigma_z$=6~cm is expected to be achieved.

\begin{figure}[tb]
\centerline{\includegraphics[width=3.5in,clip]{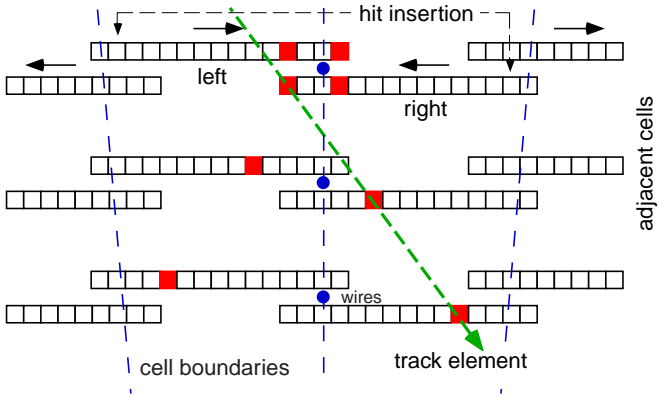}}
\caption{Hits of one cell in a trigger layer are fed into
  shift registers schematically shown in this picture. The shift register
  information is used to find track segments which are basically characterized
  by straight lines as indicated by the dashed arrow.}
\label{figure:shiftregister}
\end{figure}

For the track segment finding hits are fed into 80~MHz shift registers
implemented into the mentioned FPGAs. A first coarse track segment finding is
done by logically ORing four adjacent entries effectively reducing the
synchronization frequency to 20 MHz. For each HERA bunch crossing a
parallel search for genuine track segments is performed in all four trigger
layers and all cells consisting of three wires each.
To account for tracks crossing cell boundaries the segment finding is extended
using selected wires from neighboring cells. The main principle is shown in
figure~\ref{figure:shiftregister}. Any track is characterized by hits in the
parallel shift registers forming basically straight lines.  Left-right
ambiguities are resolved automatically when linking the track segments at the
following stage.  In order to perform the track segment finding within the
limited time given by the L1 latency, the corresponding algorithms have to be
implemented in a highly parallel and flexible way. The necessary resources are
provided by the new high density FPGAs nowadays available and their embedded CAM
functionality.

CAMs (Content Addressable Memories) can be regarded as inverse RAMs (Random
Addressable Memories) where the input patterns are compared with pre-loaded
values and matches are indicated by either signaling the corresponding
address location bit ``high'' (unencoded output) or by generating a list of
matching addresses (encoded output). They are commonly used for networking and
computing, e.g. transformation of IP addresses, data compression and as cache
tag allowing fast access to memory data.  Thus CAMs open a wide field of
applications, in particular in high energy physics where search tasks like track
linking, cluster finding and pattern recognition are common standard.
  
The integration of CAMs in FPGA devices (e.g.\ Altera \cite{cite:altera}) allows
complex high speed applications running at 100~MHz frequencies and more. As an
example the usage of tag fields is given. When operating a CAM in the so called
encoded mode a list of addresses of valid match locations is generated. 
In a second step these addresses are then used to access additional information
from a parallel RAM. This tag field method is implemented in the Fast Track
Trigger to assign track parameters to valid hit combinations.


\subsubsection{Generation of L1 Trigger Signals}
Every HERA bunch crossing a coarse track segment finding for all groups of three
wires is performed in a pipelined manner. The coarse time information
synchronized at 20~MHz and buffered in shift registers is presented in parallel
to up to 64 CAMs implemented into a single FPGA. The inputs are compared to
pre-calculated patterns of valid tracks originating from the nominal vertex.
The CAMs are running in the unencoded mode with each output bit directly
corresponding to a single bin in the $\kappa$-$\phi$ track parameter space,
where $\kappa$ represents the track curvature and $\phi$ the inclination in a
local coordinate system.

The track segment parameters from the 5 FPGAs at the front of each FEM are
collected and sent to an I/O controller. Via intermediate Merger Cards the track
segment information from 30 FEMs is transmitted to the L1~Linker Card.  For data
transmission a high speed LVDS Channel Link \cite{cite:lvds} running at 100~MHz
(600~MBytes/s) is used.

In a last processing step the track segment information from the four different
trigger layers are linked by a 3$\times$3 sliding window technique searching for
coincidences in the $\kappa$-$\phi$ track parameter space in at least 2 out of 4
trigger layers. Since most track segments can only be linked for a single bunch
crossing an event $t_0$ can be generated looking for a maximum in the number of
linked tracks. Besides such a general $t_0$ signal topological criteria and
track multiplicities as a function of different transverse momentum thresholds
may also be used to form a first level trigger decision.


\subsubsection{Refined track segment finding}
After an event has been accepted by the first trigger level the shift registers
are held and more refined track segments to be used at level 2 are extracted.
For this the coarse track segments based on the 20~MHz digitization are
serialized and expanded by restoring the original 80~MHz information. This
improves the precision in $\kappa$ and $\phi$ substantially.
In addition the $z$-position of hits stored earlier in parallel shift register
is added to obtain the full 3-dimensional track information.  Before finally
sending the refined track segments via the I/O controller to the second level
stage a further validation has to be performed to guarantee that the more
precise track parameters satisfy the criteria of a track originating from the
vertex.  This is done by checking for valid masks pre-loaded into an SRAM
serving as look-up table (LUT). If the refined segments match pre-calculated
masks a validation bit is set and the corresponding $\kappa$ and $\phi$ values
are read from a second LUT.

$\kappa$-$\phi$ values and the $z$-coordinates of the corresponding hits are
transferred via the ``Merger Cards'' to the second level of the FTT where the
individual segments are combined to global tracks.

\subsection{L2 Track Linking and Fitting} 
The level 2 stage of the FTT performs a track linking in two dimensions in
$\kappa$-$\phi$ space followed by a full 3-dimensional track fit also including
the transfered $z$-position. About 20 $\mu$s after a positive L1 decision FTT~L2
has to provide information to the central trigger logic.  In collaboration with
SCS \cite{scs} a generic multi-purpose FTT L2 board is developed which may be
used for several purposes: data merging, track linking, track fitting and
generation of trigger signals. This is realized by mounting a high density APEX
20K600E FPGA and optionally up to 4 DSPs onto the board.  Flexible I/O
configurations are realized by ``Piggy Back'' cards to interface the 100 MHz
LVDS channel links.

\begin{figure}[tb]
\centerline{\includegraphics[width=3.6in,clip]{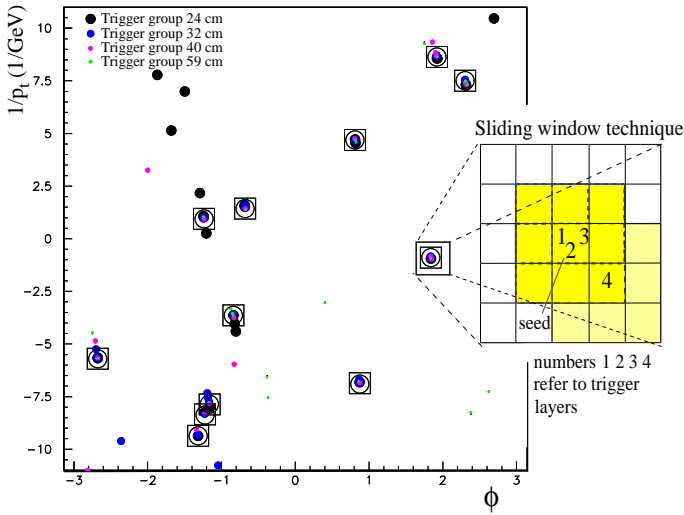}}
\caption{Schematic principle of the track linking step. The level~1
  track segment information (marked as dots) is filled into a $\kappa$-$\phi$
  histogram. Tracks successfully linked by the FTT are marked as circles while
  the boxes represent tracks found by the full H1 chamber reconstruction. The
  inset shows the principle of the 3 $\times$ 3 sliding window technique.}
\label{figure:kappa-phi}
\end{figure}

Equipped with four Piggy Back cards Merger Cards are used to bundle the data
coming from FTT level~1 onto the ``Linker Card'' where they are fed to the high
density FPGA for linking.  Track segments from the four trigger layers are
buffered in RAMs and the $\kappa$ and $\phi$ values of each layer are used to
fill 2-dimensional histograms with 40 bins in $\kappa$ and 640 bins in $\phi$.
>From the RAM a list of track seeds in the $\kappa$-$\phi$ plane is build
starting at the first trigger layer. The list of track segments is processed and
25 parallel CAMs per layer are used to find track segment clusters in a single
processing step. By a 3$\times$3 sliding window technique the exact peak
position is then determined and in case of a valid track the full track
information is restored from the parallel RAM. The basic principle of the track
linking procedure is illustrated in figure~\ref{figure:kappa-phi}.

The full track information of the linked tracks is sent to the ``Fitter Cards''
which are equipped with 4 floating point DSPs (Texas Instruments TMS320C6701
\cite{cite:dsp}). A load balancing algorithm is used to share the tracks among
different DSPs. The procedure starts with a non-iterative fit in the $r$-$\phi$
plane~\cite{cite:karimaeki} constraining the tracks to originate from the
primary vertex.
Afterwards a fit in the $r$-$z$ plane is performed, using additional information
on the $z$-position of the vertex provided by the H1 trigger system.  All
successfully fitted tracks are collected by the ``L2 Decision Card'' where track
based event quantities like momentum sums
are used to generate the final L2 trigger signals. If the event is accepted by
the second level trigger the complete set of 3-dimensional tracks is passed on
to FTT L3.

\subsection{Searching for Particle Resonances}
A farm of up to 16 single CPU VME boards is used to calculate an L3 trigger
decision.
Track information coming from FTT L2 are sent via a fast LVDS
receiver-transmitter chain to special receiver boards.  On these boards the data
are buffered and using a 20~MHz FPDP (Front Panel Data Port \cite{cite:fpdp})
sent via PMCs (PCI Mezzanine Cards \cite{cite:pmc}) to the local memory on the
CPU boards.  This mechanism ensures that all track data are available on each
board after a maximum transmission time of $<$10~$\mu$s leaving plenty of time
for sophisticated analyses.  The aim is to provide an environment where analysis
groups can easily implement their analysis code. This is guaranteed by the use
of a commercial real time operating system including standard compiler software
to allow easy implementation of e.g.\ C-code. The final L3 trigger decision will
be primarily based on the track information received from FTT L2 and a main task
will be the search for particle resonances within a high multiplicity
environment of up to 48 tracks. However, additional information from other first
level trigger systems can be made available allowing for a more sophisticated
decision based on correlations amongst different subdetectors.

\section{Simulated Performance}
The different performance studies maybe categorized into ({\it A}\,)~timing
estimates for the different algorithms to be implemented on FPGAs and DSPs done
using specialized software~\cite{cite:Quartus,cite:DSP}, and ({\it B}\,)~tests
of the FTT track reconstruction abilities based on Monte Carlo programs. For
both categories extensive simulations have been done showing that the Fast Track
Trigger fulfills the demands of the existing system and the upgraded HERA
collider.

\begin{table}[b]
\caption{FTT latencies for L1--L3. The estimated times are based on simulations
  using specialized software~\protect\cite{cite:Quartus,cite:DSP}. They have to be
  compared to the time constraints given by the H1 trigger system of 2.3~$\mu$s,
  23~$\mu$s and 100~$\mu$s for the overall L1, L2 and L3 decision time.}
\begin{center}
{\small
\begin{tabular}{clr}       \hline      
\bf \raisebox{-.5ex}{Level} & \bf \raisebox{-.5ex}{Task}  
                 &  \multicolumn{1}{c}{\bf Latency}   \\ 
          &      &  \multicolumn{1}{c}{[cumulated]}   \\ \hline
L1 & Ionisation \& drift time & 1.1\,$\mu$s \phantom{0} \\ 
& Analogue cable delays                & 1.3\,$\mu$s \phantom{0} \\
& Coarse track segment finding         & 1.5\,$\mu$s \phantom{0} \\
& Transmission to L1~Linker            & 1.9\,$\mu$s \phantom{0} \\
& Track linking                        & 2.0\,$\mu$s \phantom{0} \\ 
& Trigger signal transmission          & 2.1\,$\mu$s \phantom{0} \\
& L1 Trigger decision                  & 2.3\,$\mu$s \phantom{0} \\ \hline
L2 &  Track segment refinement         & 3.5\,$\mu$s \phantom{0} \\ 
& Transmission to L2~Linker            & 3.9\,$\mu$s \phantom{0} \\
& L2~Linking (48 tracks)               & 9.1\,$\mu$s \phantom{0} \\
& Track fitting (48 tracks)            & 12.9\,$\mu$s \phantom{0} \\ 
& Generation of trigger signals        & 21.9\,$\mu$s \phantom{0} \\ 
& L2~Trigger decision                  & 23.0\,$\mu$s \phantom{0} \\ \hline
L3 & Transmission to FTT L3            & $<$30\,$\mu$s \phantom{0} \\ 
& L3 analysis \& decision              & $<$100\,$\mu$s \phantom{0} \\ \hline 
\end{tabular}}
\end{center}
\label{tab:timing}
\end{table}

\subsection{Timing}
The FTT latencies of all three stages (L1--L3) are given in
table~\ref{tab:timing}. The numbers in the table should be compared to the time
constraints given by the H1 trigger system of 2.3~$\mu$s, 23~$\mu$s and
100~$\mu$s for the overal L1, L2 and L3 decision time, respectively. While time
requirements for FTT L1 are rather tight mainly due to the large delay which
arises from the maximum drift time to the CJC sense wires, the demands on FTT L2
are less stringent. After the track linking step the remaining time for track
based calculations of $\sim$9~$\mu$s is long enough to even allow invariant mass
sums in low multiplicity events to be completed.  For the FTT L3 farm it has
been checked that modern CPUs can cope with the combinatorial complexity of the
search for e.g. the golden decay of $D^*$ mesons ($D^* \rightarrow D^0
\pi_{slow} \rightarrow K \pi \pi_{slow}$) within the required time
$\sim$100~$\mu$s.

\begin{figure}[tb]
\vspace{.3cm}
\centerline{\includegraphics[width=3.0in,clip]{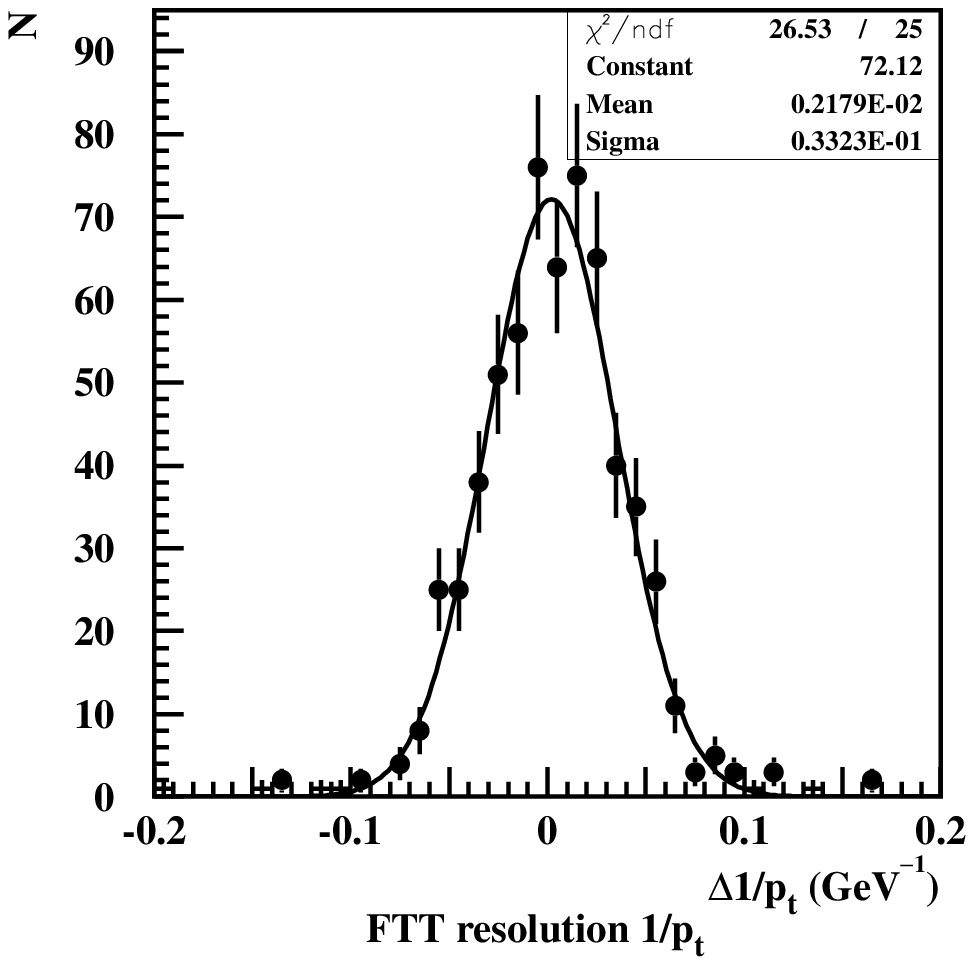}}\vspace{.4cm}
\centerline{\includegraphics[width=3.0in,clip]{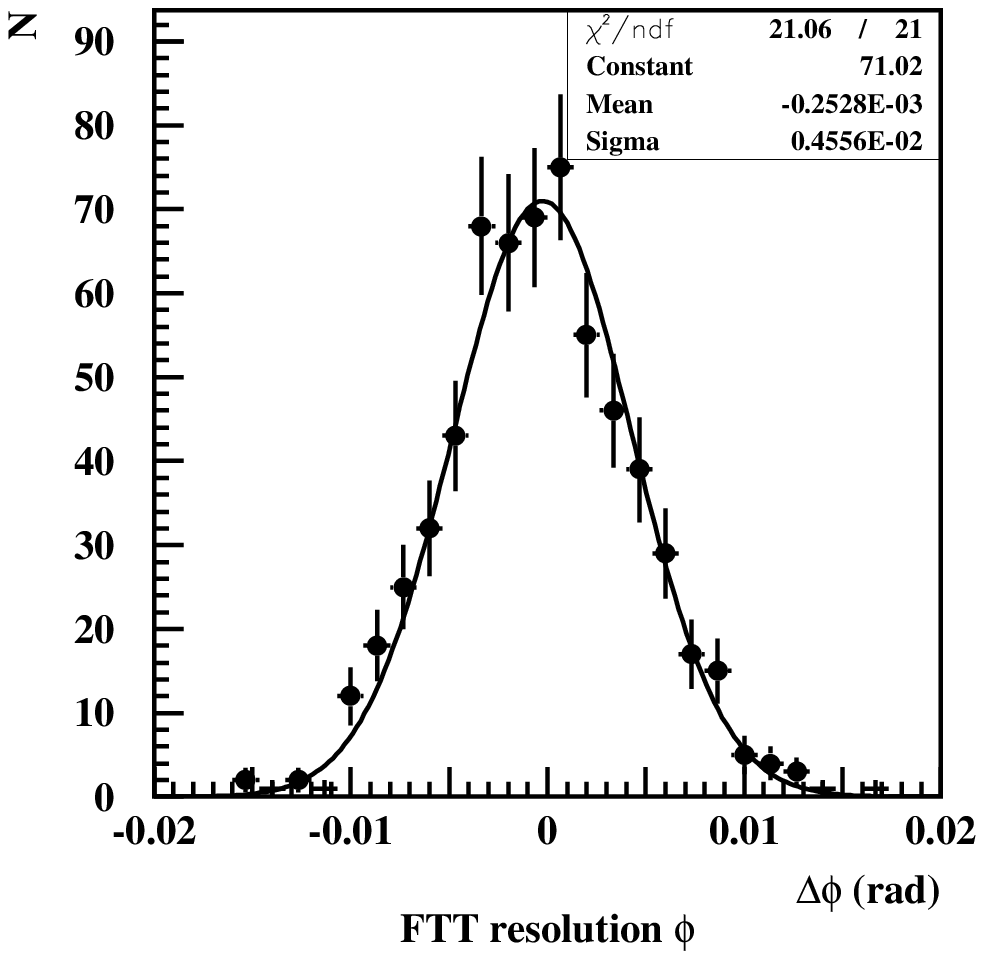}}\vspace{.2cm}
\caption{Track resolution of the simulated FTT algorithm in $1/p_t$ and $\phi$ relative 
  to the full off-line CJC reconstruction. The tracks studied are taken from a
  sample of $D^* \rightarrow K \pi \pi_{slow}$ candidates. Gaussian fits are
  shown to both distributions.}
\label{figure:resolution}
\end{figure}

\subsection{Physics Return}
In order to allow for detailed performance studies a software package has been
developed which simulates the digital part of the first, second and third level
of the FTT. It has been used to obtain track resolutions, study efficiencies and
to determine the robustness of the system concerning its dependence on the
chamber performance. In addition different implementations of the available
track finding, linking and fitting algorithms have been compared and could be
optimized.

The FTT is designed to find tracks down to a transverse momentum $p_t$=100~MeV.
The expected resolutions relative to the nominal H1 reconstruction in $1/p_t$
and $\phi$ can be extracted from figure~\ref{figure:resolution} giving
$\sigma(1/p_t) = 0.03 \mbox{ GeV}^{-1}$ and $\sigma(\phi) = 4.6$ mrad.

To investigate the full power of the Fast Trigger up to level 3 the $D^*$
channel, $D^* \rightarrow D^0 \pi_{slow} \rightarrow K \pi \pi_{slow}$,
mentioned above has been analysed in detail revealing that overall efficiencies
in excess of 80\,\% relative to a typical off-line selection can be achieved.
The corresponding trigger rates can be kept reasonably low as needed not to
incur excessive dead time at the early trigger stages.

\begin{figure}[t]
\vspace{.3cm}
\begin{center}
\epsfig{file=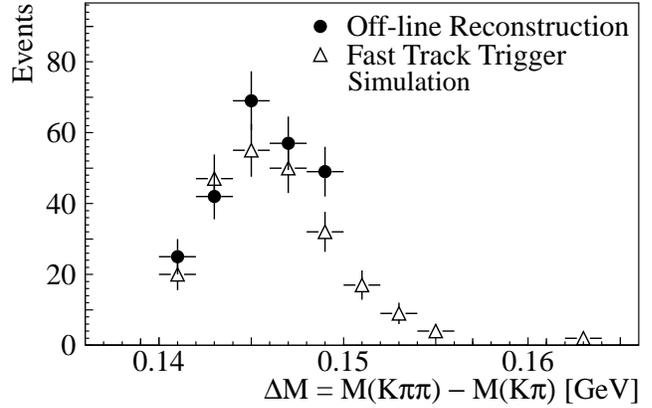,width=0.48\textwidth}
\end{center}
\caption{Illustration of the expected $\Delta m = m(K \pi \pi_{\rm 
    slow}) - m(K \pi)$ resolution of the proposed trigger.  The solid circles
  show the $\Delta m$ distribution of a sample of $D^*$ candidates from DIS
  events collected by H1 in 1997, as reconstructed using the best available
  off-line analysis tools ($|m(K\pi)-m(D^0)|<$0.08~GeV, $\Delta m<$0.15~GeV).
  The open triangles show the $\Delta m$ distribution for the same sample of
  events with tracks reconstructed using the simulation of the proposed track
  trigger.}
\label{ref:delM}
\end{figure}

\begin{figure}[tb]
\centerline{\includegraphics[width=3.0in,clip]{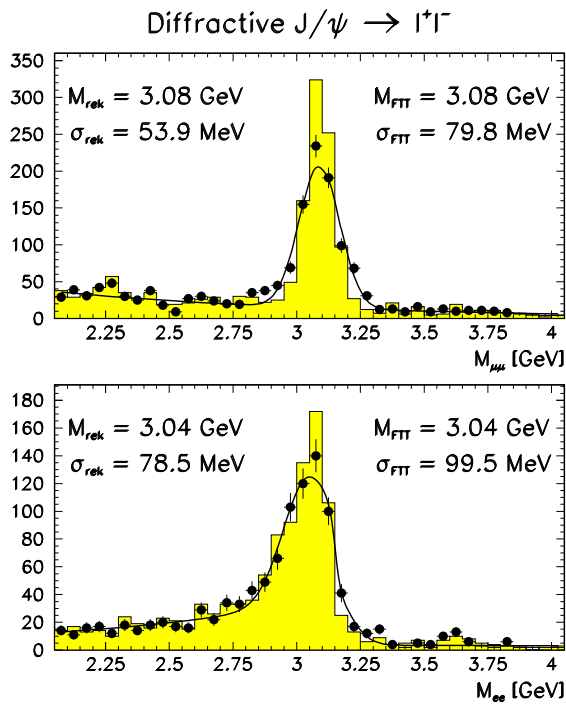}}
\caption{Reconstruction of $J/\psi$'s decaying into two leptons, top
  $J/\psi \rightarrow \mu^+ \mu^-$, bottom $J/\psi \rightarrow e^+ e^-$. The
  full histograms show the invariant mass distributions as reconstructed with
  the full readout information and the best available off-line tools. The values
  on the left hand side for the peak value and the width belong to this data
  sample. The solid data points show the distributions as reconstructed from
  tracks produced by the FTT simulation giving the peak position and width shown
  on the right hand side.}
\label{figure:jpsi}
\end{figure}

The FTT abilities up to L2 were tested by feeding low multiplicity H1 data of
$J/\psi$ meson candidates into the simulation followed by a full $J/\psi$
analysis using the FTT tracks. The outcome of the invariant $J/\psi$ mass
spectrum is compared to the one obtained using the standard H1 reconstruction
software. Figure~\ref{figure:jpsi} shows the corresponding mass spectra for
$J/\psi$'s decaying into muons and electrons convincingly showing the potential
of the FTT. Further physics channels have been analyzed in a similar way leading
to similar results.

Finally, in order to study the influence of the potentially larger background
expected after the HERA upgrade, tracks from beam-gas background have been
artificially added to genuine events.  Furthermore, a decreased single wire
efficiency, decreased $z$-resolution and additional random noise were included
in the simulation to test the stability of the system.  No substantial
performance loss has been observed for any reasonable scenario studied.

\section{Conclusion}
The H1 collaboration is in the process of building a Fast Track Trigger (FTT) to
cope with the demands of the high luminosity HERA upgrade planned for 2001. It
is based on novel technologies of integrated circuits making extensive use of
the embedded CAM functionality of modern high density FPGAs and shall supply
information to the first three trigger levels of the H1 triggering system. The
FTT will be capable of performing complex algorithms such as fast track segment
finding, track linking and fitting. Apart from simple track based quantities to
be provided after the L1 latency of 2.3~$\mu$s it will perform sophisticated
analyses, such that momentum sums and invariant mass sums can be provided at the
L2 (23~$\mu$s) or L3 (100~$\mu$s) decision time depending on event size and
complexity of the calculations. Simulations show a high physics potential and a
large selectivity of the new system.

\newpage

\nocite{*}
\bibliographystyle{IEEE}

%

\end{document}